\documentclass[aps,pre, reprint,floatfix]{revtex4-1}

\usepackage{amsfonts}
\usepackage{amsmath}
\usepackage{graphicx}
\usepackage{calligra}
\usepackage{array}
\newcommand{\D}[2]{\frac{\mathrm{d} #1}{\mathrm{d} #2}}
\newcommand{\iD}[2]{\mathrm{d}#1/\mathrm{d}#2}

\newcommand{\e}{\mathrm{e}}
\newcommand{\I}{\mathrm{i}}

\renewcommand{\mod}[1]{\left|#1\right|}

\renewcommand{\P}[2]{\frac{\partial #1}{\partial #2}}

\newcommand{\df}[1]{\mspace{2mu}  \mathrm{d}#1}
\newcommand{\reals}{\mathbb{R}}
\newcommand{\complex}{\mathbb{C}}
\DeclareMathAlphabet{\mathpzc}{OT1}{pzc}{m}{it}
\def\res{\mathop{\text{Res}}\limits}
\def\intr{\mathop{\text{int}}}
\newcommand{\conj}[1]{\overline{#1}}
\newcommand{\iip}[2]{\langle #1,#2\rangle}
\newcommand{\ip}[2]{\left\langle #1,#2\right\rangle}

\begin{document}
\title{Phase and Amplitude dynamics of nonlinearly coupled oscillators.}
\date{\today}
\author{P. Cudmore}
\author{C.A. Holmes}
\affiliation{School of Mathematical and Physical Sciences, The University of 
Queensland}

\begin{abstract}
This paper addresses the amplitude and phase dynamics of a large system non-linear coupled,
non-identical damped harmonic oscillators, which is based on recent research in coupled oscillation in optomechanics. 
Our goal is to investigate the existence and stability of collective behaviour which occurs 
due to a play-off between the distribution of individual oscillator frequency and the type of 
nonlinear coupling.
We show that this system exhibits synchronisation, where all oscillators are 
rotating at the same rate, and that in the synchronised state the system has a regular structure 
related to the distribution of the frequencies of the individual oscillators. 
Using a geometric description we show how changes in the non-linear coupling function can 
cause pitchfork and saddle-node bifurcations which create or destroy stable 
and unstable synchronised solutions. We apply these results to show how in-phase and anti-phase solutions 
are created in a system with a bi-modal distribution of frequencies.
\end{abstract}
\maketitle
\begin{quotation}
 Recent advances in cavity optomechanics highlight the significance of all-to-all coupled oscillator systems where the oscillators are linear and the
 coupling is nonlinear. In such systems, amplitude dynamics play an important role and as such exclude the use of phase oscillator models.
 Nonetheless, we show that synchronised states do exist, and their existence and stability not only depends on the distribution of the natural frequencies
 of the oscillators, but also on the mean field amplitude of the synchronised state.
 Finally, we demonstrate the importance of these results for some physically relevant coupling functions.
\end{quotation}

\section{Introduction}
Synchronisation is a well studied phenomenon in which arrays of coupled 
oscillatory systems interact and adjust their behaviour to form emergent collective motion. 
Systems which exhibit synchronisation occur in a wide variety of contexts including 
chemical oscillators\cite{kuramoto75}, biological 
systems\cite{winfree90,ermentrout90}, mechanical systems
\cite{synch,  guckenhiemer} and Josepson Junctions\cite{WatanabeStrogatz94}. 
The key question when looking at systems exhibiting synchronisation is under 
what conditions does collective behaviour appear. In most cases, the answer can 
be phrased as a trade off between the nature of the coupling and population
heterogeneity.

In the simplest model of an oscillatory systems the individual amplitudes are neglected and each 
oscillator is 
described solely by its phase $\theta_j$. Many of these phase oscillator models result from 
identically coupled limit cycle oscillators which can be shown to typically exhibit 
approximately sinusoidal coupling of each individual to all others, as initially detailed by 
Kuramoto in 1975\cite{kuramoto75}.

Significantly it has been shown\cite{strogatz00,kuramoto84} that the state of 
such a system can be described by a population `mean field' which represents the 
phase coherence of the system.
When the coupling strength is below a critical value, dependent on the 
distribution of frequencies, the oscillators act as if they were
uncoupled and the mean field approaches zero. For coupling strengths above the the critical value, the mean field 
approaches a non-zero steady state where a portion of the population 
behaves in unison. In the limit where the coupling strength is infinite the value 
of the mean field approaches one, indicating that all oscillators are identical in 
both phase and frequency.

Whilst the phase oscillator description is extremely useful, there are many 
situations where the amplitude dynamics don't decouple from the phase dynamics. 
One model which encompasses both phase and amplitude behaviour is the 
Stuart-Landau Oscillator\cite{cross2006,synch, lee2013},
\begin{equation}
  \D{z_j}{t} =(\alpha_j + \I\omega_j)z_j -(\beta_j +\I\gamma_j)\mod{z_j}^2z_j, j = 1\ldots n\label{SLOsc} 
 \end{equation}
or limit-cycle oscillator model. Eq.~\eqref{SLOsc} describes the evolution of an ensemble of $n$  
complex-valued oscillators $z_j$ with radius $|z_j|$ and phase $\arg z_j$. The real parameters $\alpha_j$, 
$\beta_j$,$\omega_j$ and $\gamma_j$ represent the linear 
and non-linear growth, natural frequency and non-linear frequency pulling respectively. 
While it is usual to assume a distribution of natural frequencies, so that $\omega_j$ is the frequency of the $j$th oscillator, 
the remaining parameters are generally assumed identical across the ensemble, with $\alpha$
and $\beta$ assumed positive in order for each $z_j$ to have a limit cycle at $z_j = \sqrt{\alpha/\beta}$.
Much work has been done\cite{Mirollo90, matthews91, synch, lee2013, cross2006} 
when these systems are coupled linearly to a mean field 
 \begin{equation}
  z = \frac{1}{n}\sum_{j=1}^nz_j. \label{eqmeanfield}
 \end{equation}
 and a comprehensive survey of the different types of behaviour can be found in 
Matthews\cite{matthews91}. 
 
Here we consider a nonlinear system of coupled oscillators where the non-linearities appear in
the coupling function. As far as we are aware, this case 
has not yet been investigated and is not covered by current results on Stuart Landau oscillators.
Our interest in this case is motivated by a physical problem where a population 
of nanoscale resonators are coupled via an electromagnetic field. The resonators 
are modelled as damped harmonic oscillators and coupling is non-linear in mean field amplitude 
only such that the coupling takes 
the form
  $zF(\mod{z})$. We will refer to $F:\reals\rightarrow 
\complex$ as coupling function 
  throughout the remainder of this work.
 
   The goal of this work is to identify the dynamic states of non-linearly 
coupled linear oscillators and 
 investigate how the heterogeneity of the system interacts with the nonlinear 
coupling to produce coherent
 behaviour. We aim to do this with generality towards the shape of the frequency 
distribution and show
 that changing the modality of the distribution can produce the in-phase and 
anti-phase solutions 
 previously investigated\cite{cathy2012}.

 The rest of this paper is organised as follows.
In Section \ref{sec:bg} we discuss a motivating example of a physical system 
and introduce the dispersion function.
In Section \ref{sec:fp} we present our main result that synchronised solutions, which
are fixed points of the system in a rotating frame, lie on a circle in the complex plane 
co-rotating with the mean field whose size, 
location and rotational velocity depend on the interaction 
between the non-linear coupling function and the dispersion function.
The stability criteria are also shown to depend upon the dispersion function.
In Sections \ref{sec:lor}-\ref{sec:bm} we discuss the effect of taking different
distribution function $g(\omega)$.
Section \ref{sec:lor} discusses how in the case of the Cauchy distribution, the fixed 
point and stability characteristics drastically simplify, depending only on the real part of the coupling function.
In Section \ref{sec:uni} we discuss how, in the more general case, the relationship between the complex 
coupling function and dispersion function determine the fixed points and stability of the system.
In section \ref{sec:bm} we apply our main results to a bi-modal frequency 
distribution to produce
in-phase and anti-phase solutions found in \cite{cathy2012}. 
Finally, a summary of our results and further discussion is found in section 
\ref{sec:conc}
 \section{Background}
 \label{sec:bg}
\subsection{Physical Motivation}
The primary motivation for our study is the recent paper by Holmes, Milburn and 
Meaney\cite{cathy2012}, which investigated the dynamics of nano-mechanical 
resonators in an microwave cavity. These systems have potential applications in 
quantum metrology\cite{qns}, quantum information science\cite{bagheri2011} and 
may provide insight into the behaviour below the quantum limit\cite{Phelps2011}. 
The physical system is comprised of an array of nano-scale mechanical resonators embedded in an 
optical cavity. When the optical cavity is externally driven, the system exhibits 
radiation pressure coupling where the displacement of each resonator modulates the 
resonant frequency of the common electromagnetic field, which in turn forces the mechanical mode of
each resonator. 
This interaction mediates a highly non-linear all-to-all coupling between each resonator. 

When all the oscillators are identical the synchronised state is dominated by oscillatory 
solutions created by Hopf bifurcations and 
pairs of saddle-node bifurcations of periodic orbits (Figure 
2\cite{cathy2012}).  
The bifurcation parameters are functions of magnitude and frequency of the 
external forcing. Changes in these parameters effectively change the shape and 
structure of the coupling function and as such it is useful to treat the 
coupling function with some generality.

When two or three identical population groups are introduced, the system 
exhibits states defined as in-phase synchronised solutions (oscillator angles $\theta_1 \approx \theta_2$ and 
rotating at same rate) and 
anti-phase synchronised solutions (oscillator angles $\theta_1 \approx \theta_2+\pi$ 
and rotating at same rate). 
But as the frequency separation is varied the oscillatory solutions 
differ in amplitude 
such that the sum of population amplitudes $r_1+r_2$ is approximately constant for the 
in-phase solution and the difference
$r_1-r_2$ is approximately constant for anti-phase solutions. The desire to 
understand this behaviour in the larger context of distributions of frequencies 
leads us to consider bimodal as well as unimodal distributions.
\subsection{Model}
The mechanical resonators are described as damped harmonic oscillators and as such 
we neglect the non-linear damping and frequency pulling terms in 
Eq.~\eqref{SLOsc}
($\beta$,$\gamma=0$). 
We assume that the damping rate is identical for each oscillator and under 
correct normalisation we can set 
$\alpha=-1$. 
Each natural frequency $\omega_j$ is randomly sampled from some probability 
distribution 
with density function $g(\omega)$ which, without loss of generality, we can assume to have zero mean (if the distribution has a mean 
at $\omega =\bar{\omega}$, a change of variables $z' = z\e^{\I\bar{\omega}t}$, 
$\omega'=\omega - \bar{\omega}$ will 
shift the mean to zero). 
The state of the system is thus $2n$ dimensional in $(z_j,\conj{z_j})$ with the equations of motion given by
 \begin{equation}
 \D{z_j}{t} = -(1 - \I\omega_j)z_i + zF(\mod{z}),\label{nrampeqn},
 \end{equation}
 and the complex conjugate, with the mean field $z$ as per Eq.~\eqref{eqmeanfield}.

We define the mean field amplitude 
$ r = \mod{z}$, phase $\Theta = \arg z$ and 
oscillation frequency 
$ \Omega = \iD{\Theta}{t}$.
Moving to a frame of reference co-rotating with $\Theta$ (by a change variables 
$z_j\rightarrow z_j\e^{\I\Theta}$)
results in the following system of equations for the individual oscillators
\begin{subequations}
 \begin{equation}
 \D{z_j}{t} = -[1 + \I(\Omega - \omega_j)]z_j + rF(r),\label{ampeqn}
 \end{equation}
\begin{equation}
 r = \frac{1}{n}\sum_{j=1}^n z_j.\label{meanfield}
\end{equation}
\end{subequations}

In the decoupled system, when $F(r)\equiv 0$, each oscillator has a stable fixed point at the 
origin. We are therefore interested in how coupling forces individuals off 
this stable fixed point. We consider the state with $z_j=0$ for all $j$ akin to 
the incoherent state of the Kuramoto model in the sense that there is 
insufficient coupling to produce macroscopic behaviour so oscillators follow 
their inherent behaviour and exponentially decay.

Broadly speaking, the transition out of this state occurs when there is sufficiently strong linear 
coupling ($F(0)$) with a thin enough distribution of frequencies to destabilise the origin. 
The actual details of this interaction between the strength of the coupling and 
the spread of the distribution of frequencies $g(\omega)$ depends on the dispersion function
\begin{equation}
    f(\mu) = \int_{-\infty}^\infty\frac{g(\omega)}{\mu-\I\omega}\df{\omega} 
\qquad \Re[\mu]>0\label{fmu}.
\end{equation}
This is not surprising as this dispersion function has previously appeared in 
literature on coupled Stuart-Landau oscillators 
\cite{Mirollo90,matthews91,lee2013} as the limit as $n\rightarrow \infty$ of
\begin{equation}
 f_n(\mu) = \frac{1}{n}\sum_{j=1}^n \frac{1}{\mu-\I\omega_j}. \label{fmun}
\end{equation}

In fact amongst other properties of $f$, we will use the results previously shown by Mirollo and Strogatz\cite{Mirollo90} on
the relation between solutions of $f_n(\mu) = K$ for finite $n$ and $f(\mu) = k$ to obtain existence and 
stability criteria for the synchronised state.

\section{Fixed States and stability}\label{sec:fp}
A simplifying feature of this system is the equivalence
between steady state fixed points of the the mean field $(r,\Omega)$, and steady state fixed points of 
individual oscillators $z_j$. This is because the decoupled system is linear and when decoupled each oscillator
will relax to the equilibrium state.
To see this, we consider the fixed points of the $j$th 
oscillator Eq.~\eqref{ampeqn} which occur at
\begin{equation}
 z_j = \frac{rF(r)}{1+\I(\Omega -\omega_j)}. \label{fixedpoint}
\end{equation}
Suppose $z_j$ is fixed for some $j$, then by rearranging Eq.~\eqref{fixedpoint} 
we have for some 
$k\ne j$,
\[\D{z_j}{t} = -[1+\I(\Omega -\omega_k)]z_k + [1+\I(\Omega -\omega_j)]z_j.\] 
As this is now a linear differential equation, $z_k$ decays to a solution 
\[z_k = \frac{1+\I(\Omega -\omega_j)}{1+\I(\Omega -\omega_k)}z_j.\] 
This implies that $z_j$ is constant for all 
$j$ and also from Eq.~\eqref{meanfield} that $r$ is constant.

Conversely, if we suppose the mean field $r$ is constant, Eq.~\eqref{ampeqn} can 
be integrated to show that each $z_j$ approaches Eq.~\eqref{fixedpoint}.

In this sense synchronised solutions appear as the fixed points of 
Eq.~\eqref{ampeqn} and are completely characterised by the fixed points in 
$(r,\Omega)$.
In the synchronised state, Eq.~\eqref{fixedpoint} gives the amplitude and phase of the $j$th 
oscillator as a function of the mean field amplitude and frequency, and the oscillators
natural frequency $\omega_j$. For a fixed $(r,\Omega)$, we can interpret Eq.~\eqref{fixedpoint} as a linear
fractional transform which maps a point $\omega_j$ on real line to a point $z_j$ in the complex plane. This can be 
understood by noting that the image of a vertical line with real part one under the
conformal map $\xi \rightarrow 1/\xi$ produces a circle of diameter one in the the right half 
plane intersecting the points zero and one.

\begin{figure}
 \includegraphics{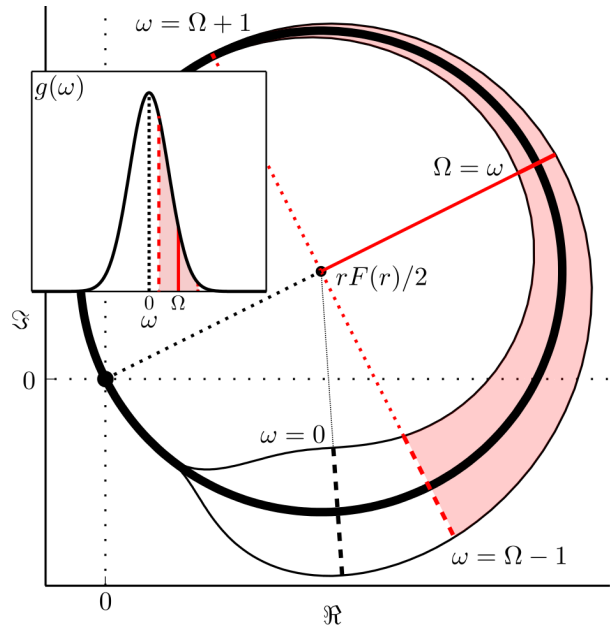}
 \caption{\label{fig:fixedpoint}
In the synchronised state, all oscillators are at a fixed position on a circle 
in the complex plane $z$, co-rotating with the mean field. The shaded and unshaded 
area provides a graphical representation of how the density of frequencies $g$  
maps to a density of position on the circle. Oscillators with frequencies 
between $(\Omega-1,\Omega+1)$ end up on the arc furthest from the origin and 
thus have the largest amplitude.}
\end{figure}

Thus the fixed points all lie on the circle with
radius $|rF(r)|/2$ and centred at $rF(r)/2$ as shown in Figure~\ref{fig:fixedpoint}, 
where the oscillators with frequencies $\omega_i\in(\Omega-1,\Omega+1)$ are mapped to the arc 
furthest from the origin.

\subsection{Fixed points of $r,\Omega$}
We now consider the dynamics of the mean field in order to characterise 
the existence and stability of the
synchronised solution. The equation governing the time evolution of the mean field $r$ 
can be found by differentiating Eq.~\eqref{meanfield} to produce
\begin{equation}
 \D{r}{t} = -r(1+\I\Omega) +rF(r) - \frac{\I}{n}\sum_{j=1}^n\omega_jz_j 
\label{eqmfde}.
\end{equation}
The fixed points of Eq.~\eqref{eqmfde} occur when each $z_j$ is also fixed by Eq. 
\eqref{fixedpoint} and $r$ satisfies
\[
  0 = r(1+\I\Omega)\left[1 - 
\frac{F(r)}{n}\sum_j\frac{1}{(1+\I\Omega)-\I\omega_j}\right].
\]
We consider the trivial solution
\[  r =0,  \]
 akin to the incoherent state in the Kuramoto model as it represents a state where the coupling 
has not been sufficient to counteract the natural behaviour of the origin, i.e, to decay to zero. 

The remaining branch
\[
  F(r) = \left(\frac{1}{n}\sum_{j=1}^n\frac{1}{(1+\I\Omega) - 
\I\omega_j}\right)^{-1},  
\]
 describes the fixed point in terms of an interaction between the non-linear 
coupling function of the 
 mean field amplitude, and the dispersion of frequencies $\omega_i$ around the 
dominant mode 
$\Omega$. 
For sufficiently large $n$, this is well approximated by 
\begin{equation}
 F(r)  = f(1+\I\Omega)^{-1}  \label{eqfp}.
\end{equation}
This result follows from Theorem 
1\cite{Mirollo90}
so that if we fix some $K$ in the range of $f_n(\mu)$, then in the limit 
 as $n\rightarrow \infty$, $f_n(\mu) = K^{-1}$ and $f(\mu)=K^{-1}$ almost surely have the 
same bounded number of solutions, independent of $n$, in any vertical strip in 
the right half plane and away from the imaginary axis. 
Consequently for large $n$ the fixed points are the values of $(r,\Omega)$ which satisfy equation~\eqref{eqfp}.

\subsection{Stability}
We begin the stability calculation with the full $2n$ dimensional system in the non-rotating 
frame of reference; Eq.~\eqref{nrampeqn} with it's complex conjugate. 
Suppose that a synchronised solution exist for a particular value of $(r,\Omega)$ and thus satisfies
Eq.~\eqref{eqfp}. 
By moving to a frame of reference constantly rotating at $\Omega$; $z_j\rightarrow z_j\exp(-\I\Omega t)$,
 \begin{equation}
 z_j' = -(1+\I(\Omega-\omega_j))z_j + zF(\mod{z}),
 \end{equation}
we can linearise the system about the synchronised solution.
Using the following partials, evaluated at the fixed point when $z=\conj{z} = r$ is real
\begin{eqnarray*}
 a =\P{}{z_j}zF(\mod{z}) &=& \frac{1}{n}\left(F(r) + \frac{r}{2}\D{F}{r}\right),\\
  b= \P{}{\conj{z}_j}zF(\mod{z}) &=& \frac{r}{2n}\D{F}{r},
 \end{eqnarray*}
letting $U$ be an $n\times n$ matrix of ones and $G$ be a diagonal matrix with entries $G_{jj} = -1+\I(\omega_j-\Omega)$,
the linearised system has the following Jacobian
\[
 J = \left(\begin{matrix}
      G + aU & bU\\
      \conj{b}U &\conj{G}+ \conj{a}U
     \end{matrix}\right).
\]
The characteristic equation $\phi$, as derived in appendix \ref{appendix2}, can be factored into 
$\phi(\lambda)=\phi_l(\lambda)\phi_d(\lambda)/(f_n(1+\lambda+\I\Omega)\conj{f}_n(1+\conj{\lambda}+\I\Omega))$
where the denominator $f_n(1+\lambda+\I\Omega)\conj{f}_n(1+\conj{\lambda}+\I\Omega)$ is always bounded for $\lambda\ne -1$ and
\begin{equation*}
\phi_l(\lambda) = \prod_{j=1}^n \left[(\lambda+1)^2+(\omega_i-\Omega)^2\right], 
\end{equation*}
has zeros occurring as conjugate pairs on a vertical line in the complex plane with real part $\Re[\lambda] = -1$. As the real part of the solutions to $\phi_l(\lambda) =0$ are 
independent of both the frequency distribution and the coupling function, $\phi_l$ does not contribute to any stability changes, except to provide a strong inherent stability.
The remaining factor $\phi_d(\lambda$) controls the stability of the synchronised state
and is expressed in terms of the finite dispersion function Eq.~\eqref{fmun}, the coupling function $F$ and it's derivative. 
As per section \ref{sec:fp}, we can approximate $f_n$ with $f$, 
so finally the stability of the synchronised solution is determined by the roots of
 \begin{eqnarray}
 \phi_d &=&
 \left[\frac{1}{f(1+\lambda+\I\Omega)}-\left(F(r) + \frac{r}{2}\D{F}{r}\right)\right]\notag\\
 &&\quad\times\conj{\left[\frac{1}{f(1+\conj{\lambda}+\I\Omega)}-\left(F(r) + \frac{r}{2}\D{F}{r}\right)\right]}\notag\\
 &&\qquad-\frac{r^2}{4}\mod{\D{F}{r}}^2  \label{eqev}
\end{eqnarray}
Note that this equation is real, so that complex solutions always appear in conjugate pairs and that in the case of the zero solution, 
when $(r,\Omega)=(0,0)$, it can be simplified to
\begin{equation}
 F(0) = \frac{1}{f(1+\lambda)} \label{eq:zeroev}.
\end{equation}
 
\section{Cauchy Distributed Frequencies}\label{sec:lor}
\begin{figure}
\includegraphics{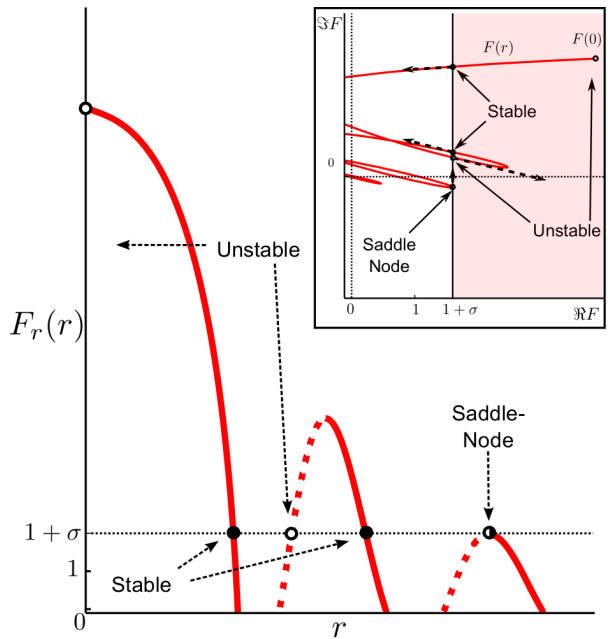} 
\caption{\label{fig:Cbp}
Fixed points and stability for the Cauchy distribution can be determined by looking at the graph of $(r,F_r(r))$ (main figure). The origin is 
an unstable fixed point as $F_r(0)>1+\sigma$. In $F$ space, this corresponds to $F(0)$ to the right of the line $1+\sigma+\I\Omega$, in the 
shaded region. 
Synchronised solutions occur when $F_r(r) = 1+\sigma$ and are stable when $\iD{F_r}{r}<0$ (on the solid part of the line) and unstable when 
$\iD{F_r}{r}>0$ (on the dotted part).
In $F$ space (inset), these solutions occur at intersections of the line $F(r)$ and the line $1+\sigma+\I\Omega$ and are stable if the tangent 
vector of $F$ points to the left.
The example function occurs in parameter set $(\epsilon,\delta)_1$ of the optomechanical coupling function derived in the 
appendix.
In this example increasing the spread $\sigma$ of the distribution, (alternatively scaling up 
$F$) will cause pairs of stable and unstable fixed points collapse in saddle 
node bifurcations of periodic orbits causing solutions to relax to the pervasive 
lower amplitude solution. 
At the bifurcation point, both $F$ and $1+\sigma +\I\omega$ will be tangential and thus purely imaginary.
The graph has been edited for visual clarity as the original graph has the value of $F_r(0)$ at a much larger magnitude putting 
it far outside the axis and as such we have scaled down that section of the curve. 
Similarly, we have radially scaled down the corresponding segment of the line $F(r)$ in the complex space (inset) so that the uppermost part of the the 
line, starting at $F(0)$ is in the visible region.}
\end{figure}

In order to investigate the effect of differing spread we consider as a first example 
the Cauchy distribution $g(\omega)$;
\[
 g(\omega) = \frac{\sigma}{\pi(\omega^2 + \sigma^2)} = 
\frac{1}{2\pi\I}\left[\frac{1}{\omega- \I\sigma} + \frac{1}{\omega+ 
\I\sigma}\right].
\]
This has a particularly simple dispersion function, as Eq.~\eqref{fmu} can be directly integrated 
for $\Re[\mu]>0$ by noting that the only pole in the upper half plane is at 
$\omega = \I\sigma$, and the integrand behaves like $\mod{\omega}^{-3}$ for 
large $\omega$. 
Using contour integration and calculating the residue from the pole at $\omega=\I\sigma$ gives
\[
 f(\mu) = \frac{1}{\mu+\sigma}.
\]

This means that the non-trivial fixed points Eq~\eqref{fixedpoint} are given by
\[
 F_r(r) = 1+\sigma, \qquad F_i(r) = \Omega,
\]
and their existence depends only on the real part of the coupling function. 

A useful way to look at the problem of existence is to consider $F(r)$ and $1/f(1+\I\Omega)$ as 
two curves in the complex plane parameterised by $r$ and $\Omega$ respectively (See inset Figure~\ref{fig:Cbp}). 
For the Cauchy distribution $1/f$ appears as a vertical line with real part 
$1+\sigma$. 
Synchronised solutions occur when these two curves intersect.

The trivial solution, which always exists, has eigenvalues given by
$\lambda =-1-\sigma + F(0)$ and its complex conjugate.
So if $F_r(0)<1+\sigma$ (inside the shaded region of Figure~\ref{fig:Cbp}), the zero 
solution is stable. 
When $F_r(0)=1+\sigma$, the origin will lose(or gain) stability via a
supercritical(subcritical) Hopf provided $\iD{F_r}{r} <0$ ($>0$). 
Figure~\ref{fig:Cbp} shows a case where the origin is unstable and there are two stable
synchronised solutions, one unstable synchronised solution, and one neutrally stable synchronised solution.

For $F(r) = 1+\sigma +\I\Omega$, Eq.\eqref{eqev}
has solutions $\lambda =0,\lambda = \D{F_r}{r}$ so that the synchronised solution is stable if $\iD{F_r}{r} <0$. 
If we continuously deform $F$ such that at the point of intersection $\iD{F_r}{r} =0$, the systems may undergo a saddle-node 
(as in Figure~\ref{fig:Cbp}) or pitchfork bifurcation of periodic orbits (depending on whether $F_r(r)$ is a local extremum or a 
point of inflection) creating or destroying synchronised solutions in the process.
As a consequence of the fact that both eigenvalues are real, it is not possible for the solution to undergo a Hopf bifurcation
to a torus.
\section{Uni-modal Distributions}\label{sec:uni}
\begin{table}
\centering
\begin{tabular}{m{1.75cm} m{3.25cm} m{3cm}}
 Distribution & \[g(\omega)\] & \[f(\mu)\]\\
\hline
 Cauchy & \[\frac{\sigma}{\pi(\sigma^2 + \omega^2)}\] & \[ \frac{1}{\sigma + 
\mu}\] \\
 Uniform & \[\begin{cases}
  \frac{1}{\pi\sigma} & \mod{\omega} < \frac{\pi}{2}\sigma\\
	0 &\text{elsewhere}\end{cases}\]
 & \[\frac{\I}{\pi\sigma}\log\left(\frac{\mu-\I\pi\sigma/2}{\mu+\I\pi\sigma/2}\right)\]\\
 Gaussian &
\[\frac{1}{\pi\sigma}\exp[-\omega^2/(\pi\sigma^2)]\] & 
\[\frac{1}{\sigma}\,\text{w}\!\left(\frac{\I\mu}{\sqrt{\pi}\sigma}\right)\]
\end{tabular}
\caption{\label{Tfmu}$f(\mu)$ for various distributions with the same modal height $g(0)$ and spread
measured by $\sigma = [\pi g(0)]^{-1}$. The 
solution for the Gaussian is
described in terms of the plasma dispersion function\cite{Olver:2010:NHMF} $\text{w}(z) = 
\exp(-z^2)\text{erfc}(-\I z)$.}
\end{table}

In the Cauchy distribution the amplitude and stability of the synchronised 
solutions only depend on
the real part of $F$, so that the stability boundary is a vertical straight line. This simplifies
the types of bifurcations that are possible, but is atypical.
We shall thus consider two symmetric unimodal distributions, the uniform and the Gaussian distributions, for 
which this property no longer holds. 
Closed forms of $f(\mu)$ for the uniform and Gaussian distributions can be found in Table \ref{Tfmu}.
As per \cite{Mirollo90,matthews91}, it is convenient to use $\sigma = [\pi g(0)]^{-1}$ as a measure of spread as this is well defined even when the variance is not.
We can note that all three unimodal distributions satisfy a scaling relation $f(\mu;\sigma) = f(\mu/\sigma;1)/\sigma$. Indeed
it is trivial to show that this holds for any continuous distribution with mean zero and spread $\sigma$ such that 
$g(\omega;\sigma) =g(\omega/\sigma;1)/\sigma$ hence we will assume $\sigma =1$ for the remainder of this section.

\begin{figure}
\includegraphics{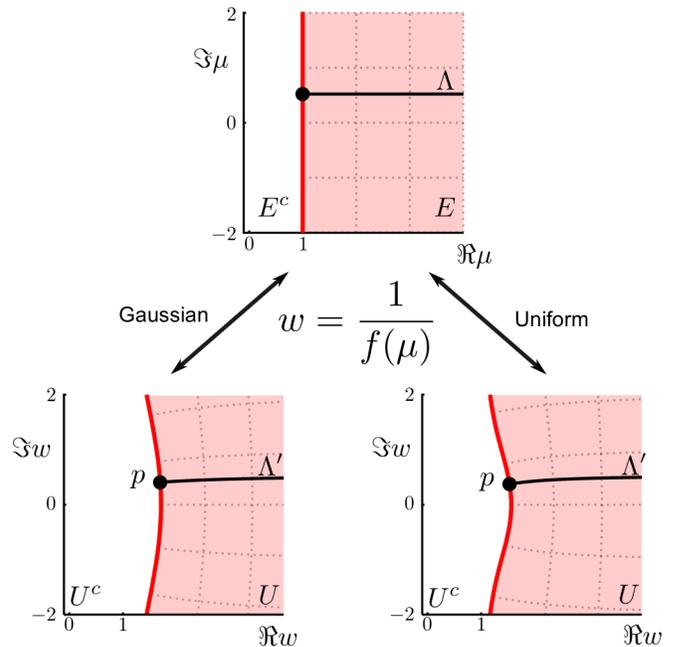}
 \caption{\label{fig:mapping}The set $E$(top, shaded region) and it images $U= 1/f(E)$ for the Gaussian (bottom left) and Uniform (bottom right) distributions with the 
 same spread $\sigma =1$. The boundary of these sets, $\partial E$ and $\partial U$ respectively, are shown as the thick lighter coloured lines. 
 The thick dark line $\Lambda'$ in the 
 bottom plots orthogonally intersects $\partial U$ at the point $p$.}
\end{figure}

In order to generalise the results in section~\ref{sec:lor}, we must first understand how $1/f(\mu)$ behaves.
In the case that $g$ is symmetric and unimodal, $f:H \rightarrow H'$ is a biholomorphic function from the right half-plane $H$ 
to it's image $H'=f(H)$ (appendix~\ref{appendix3}). The mapping $z\rightarrow 1/z$ is also biholomorphic on the right half plane.
This allows us to partition the domain of $1/f$ into two simply connected sets; $E = \{\mu:\ \Re[\mu] >1\}$, and it's open 
complement $E^c$ (Figure ~\ref{fig:mapping}, topmost graph) separated by the line $\partial E = \{\mu:\ \mu = 1+\I\Omega,\ \Omega \in \reals\}$.
The images of $E$ and $E^c$ under the biholomorphic map $1/f$ are $U$ and $U^c$ respectively 
and are bounded by $\partial E$ and $\partial U$ respectively so that if $1/f(\mu) \in U$ then $\Re[\mu]>1$. Figure~\ref{fig:mapping} shows these sets for both 
Uniform and Gaussian distributions.
As in section~\ref{sec:lor}, we consider the coupling function, $F(r)$, as a curve parametrically traced out by $r \in [0,\infty)$
and show that the intersections of $F$ and the boundary $\partial U$ correspond to synchronised solutions of the original system,
and that the stability of these solutions depend on $\iD{F}{r}$ and the set $U$.
In particular we show that the trivial solution is unstable if $F(0)\in U$ and that
a synchronised solution is unstable if the inner product of the tangent vectors of $1/f$ and $F$ at the point of intersection is positive.

\subsection{Trivial Solution}
\begin{figure}
 \includegraphics{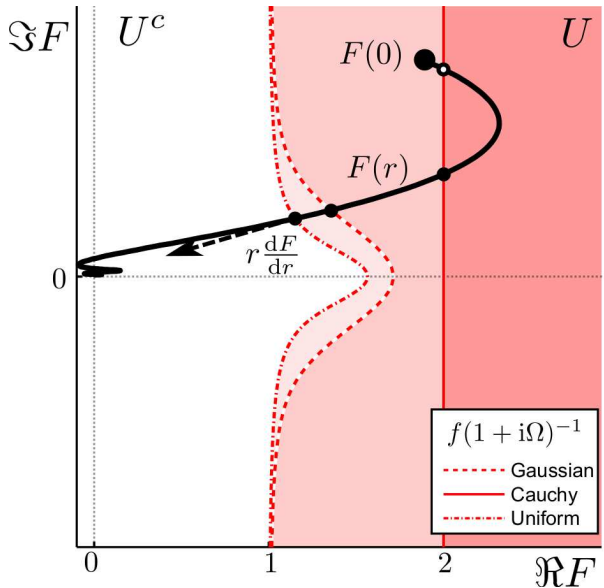}
 \caption{\label{fig:sets}For each distribution, the curve $1/f(1+\I\Omega)$ 
 describes the boundary of the shaded region $U$ where the origin is unstable. The dark 
line is the parameterisation of a physically relevant coupling function $F$, 
specified by set $(\delta,\epsilon)_3$, parameterised by the mean field amplitude $0 \le r < \infty$. 
In this example, the origin is stable for a system with Cauchy distributed frequencies (with spread $\sigma =1$)
but unstable for Uniformly and Gaussian distributed frequencies with $\sigma =1$. 
All three distributions have synchronised solutions occurring at intersections between $F$ and $\partial U$. 
The Gaussian and Uniform have one stable synchronised solution, where as the Cauchy system has an additional 
unstable synchronised solution.}
\end{figure}

The trivial solution $r=0, \Omega = 0$, which always exists, has eigenvalues $\lambda,\conj{\lambda}$ given by 
Eq.~\eqref{eq:zeroev}. We can determine whether the eigenvalues are positive or negative based on the position of $F(0)$ relative to the set $U$.
Suppose that $F(0) \in U$ then, by conformal equivalence of $U$ and $E$, $\Re[\lambda] >0$ and the origin is an unstable fixed
point. Conversely, suppose $F(0) \in U^c$, then the origin is stable.
Figure~\ref{fig:sets} shows an example where the trivial solution is stable for the Cauchy distribution,
but unstable for both the Gaussian and Uniform.

The stability of the trivial solution can change if the coupling function $F$ is continuously deformed, for example as a result of varying the strength of the coupling.
Consider Figure~\ref{fig:bifu} a); as we deform $F$ such that $F(0)$ passes from $U$, 
into $U^c$ (or vice versa) then the trivial solution gains(loses) stability at $F(0) \in \partial U$ via a subcritical (supercritical) Hopf bifurcation. 

\subsection{Synchronised Solutions}
The synchronised solutions, which satisfy equation~\eqref{eqfp}, occur at points of intersection $p(r,\Omega)$ between the line $F$ and 
the boundary $\partial E$. Figure~\ref{fig:sets} shows an example with two intersections, and thus two synchronised solutions, 
for Cauchy distributed frequencies. Only one intersection, and thus one synchronised solution occurs for the same system with either 
Gaussian or Uniformly distributed frequencies.
The stability of these synchronised solutions is less obvious and thus we present a geometric
interpretation of the roots of Eq.~\eqref{eqev} to proceed. We will now show that the stability depends the inner product of the tangent vectors of $F$ and $1/f$.

Let $P$ be the set of intersection points; $P = \partial E\cap F$ then, since $1/f$ is a conformal map and 
$F(r)$ is continuous, each $p\in P$ uniquely specifies a pair $(r,\Omega)$. 

Recall that two points $\zeta_1,\zeta_2\in\complex$ are inversions with respect to a circle with radius $r$, 
centred at $c$ if $r^2 = (\zeta_1-c)\conj{(\zeta_2-c)}$.
If we compare this to Eq.~\eqref{eqev}, then $\lambda$ is an eigenvalue for the solution at point $p$
if $\zeta_1 = f(1+\lambda+\I\Omega)^{-1}$, and $\zeta_2 = f(1+\conj{\lambda}+\I\Omega)^{-1}$ are inversions with respect to 
the circle 
\[
S_p = \left\{\zeta: \mod{p +\frac{r}{2}\D{F}{r} - \zeta} = \mod{\frac{r}{2}\D{F}{r}}\right\}.
\]
The real eigenvalues will occur on the circumference of $S_p$ as $\lambda = \conj{\lambda}$ implies that $\zeta_1=\zeta_2$.
Indeed there is always one solution $\lambda =0$ associated with the rotational frame of reference, which occurs where the circle
$S_p$ intersects the point $p$. 

\begin{figure}
\includegraphics{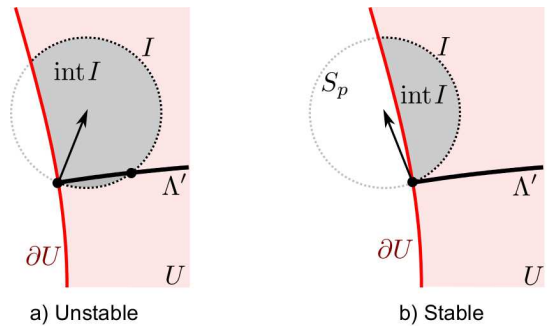}
 \caption{\label{fig:circline}
 A synchronised solution at the point $p$ has positive real eigenvalues if the line $\Lambda'$ intersects the boundary of $I$ in
 $U$ at a point other than $p$.
 In a) there is such an intersection implying the existence of at least one positive real eigenvalue for the synchronised solution.
 In b) there is no such intersection and thus the synchronised solution is stable.}
\end{figure}

Positive real eigenvalues can be identified by introducing a horizontal line 
$\Lambda = \{z:1+\lambda+\I\Omega,\ \lambda \in [0,\infty)\}$ for $\Omega$ at the point $p$ (see Figure~\ref{fig:mapping}, topmost pane). 
Clearly $\Lambda$ and $\partial E$ intersect orthogonally, so the image of this line $\Lambda' = 1/f(\Lambda)$ 
will intersect the boundary $\partial U$ orthogonally at the point $p$.
Additional real eigenvalue occur when the line $\Lambda'$ intersects the circle $S_p$.
Furthermore, if these intersections of $\Lambda'$ and $S_p$ occur in the set $U$, we know that such intersections correspond to positive real
eigenvalues, see figure~\ref{fig:circline}a.

However intersections of $\Lambda'$ and $S_p$ only occur when a section of $\Lambda'$ lies inside the circle $S_p$. 
If we let $\intr S_p$ be the disk bounded by the circle $S_p$ and $\intr I = U\cap \intr S_p$ be the subset of the disk inside the
region $U$. We shall call $I$ the union of $\intr I$ and it's boundary $\partial I$, with $p\in I$ (see figure ~\ref{fig:circline} a).
Suppose that there exists some segment of $\Lambda'$ which is inside $I$ such that $\Lambda'\cap \intr I \ne \emptyset$ then 
there must exists at least two points at which $\Lambda'$ intersects $\partial I$ and hence 
the characteristic equation for the solution at $p$ must have two real eigenvalues $\lambda\ge0$.
If there is a segment of $\Lambda'$ inside $I$ we are guaranteed that at some point $\Lambda'$ leaves $I$. 
$\Lambda'$ can't go back through the boundary, as this would require $1/f$ to no longer be bijective. 
Neither can it stay inside $I$, as the real part of $\Lambda'$ is an increasing function (Appendix~\ref{appendix3}).

There is always at least one point where $\Lambda'$ and $\partial I$ intersect; at the point $p$, where $\lambda = 0$.
Thus a sufficient condition for positive real eigenvalues is obtained by requiring that the outward normal to $\partial U $ 
at $p$ point into $\intr I$, so that $\Lambda'$ starts at $p$ and continues into the set $I$
Hence we require the inner product between the outward normal to $\partial U$ and the direction $\iD{F}{r}$ from the intersection $p$ to the centre of the circle $S_p$
to be positive.

Thus for any solution $p = p(r,\Omega)$, if
\begin{equation}
  \ip{\D{F}{r}\bigg|_r,}{\D{}{\mu}\left(\frac{1}{f}\right)\bigg|_{1+\I\Omega}} >0 \label{unimodalstab}
\end{equation}
then Eq.~\eqref{eqev} has a positive real eigenvalue and the synchronised solution is unstable, where $\iip{u}{v} = \Re[ u\conj{v}]$ is 
the geometric inner product. 

Suppose that $\Lambda' \cap I = \{p\}$, then the synchronised solution at the point $p$ is stable. 
Clearly a necessary condition for this to occur is the converse of Eq.~\eqref{unimodalstab} that $\Lambda'$ doesn't immediately enter $\mathrm{int}\ I$. 

Numerical investigation suggests that for symmetric unimodal distributions, there are bounds on how curved the line $\Lambda'$
is allowed to become and indeed it seems that as $\lambda$ increases, $\Lambda'$ asymptotes to the horizontal.
This leads us to think that Eq.~\eqref{unimodalstab} is both necessary and sufficient for a synchronised solution at point
$p$ to be unstable, as bounds on the curvature of $\Lambda'$ would
disallow it from curling back upon itself enough to intersect the circle $S_p$ a second time (or third time, if Eq.~\eqref{unimodalstab} holds).

Eq.~\eqref{eqev} allows for conjugate pairs of eigenvalues when $\zeta_1= f(1+\lambda+\I\Omega)^{-1}$ and 
$\zeta_2=f(1+\conj{\lambda}+\I\Omega)^{-1}$ are inversions with respect to the circle $S_p$. These do not appear in the case of 
Cauchy distributed frequencies so can be though of as a feature of the curvature of mapping $1/f$. We conjecture that if conjugate pairs do 
appear, they must have a real part less than the larger real eigenvalue, and thus do not contribute to the overall 
stability of a synchronised solution.

\subsection{Bifurcations of Synchronised Solutions}
\begin{figure}
 \includegraphics{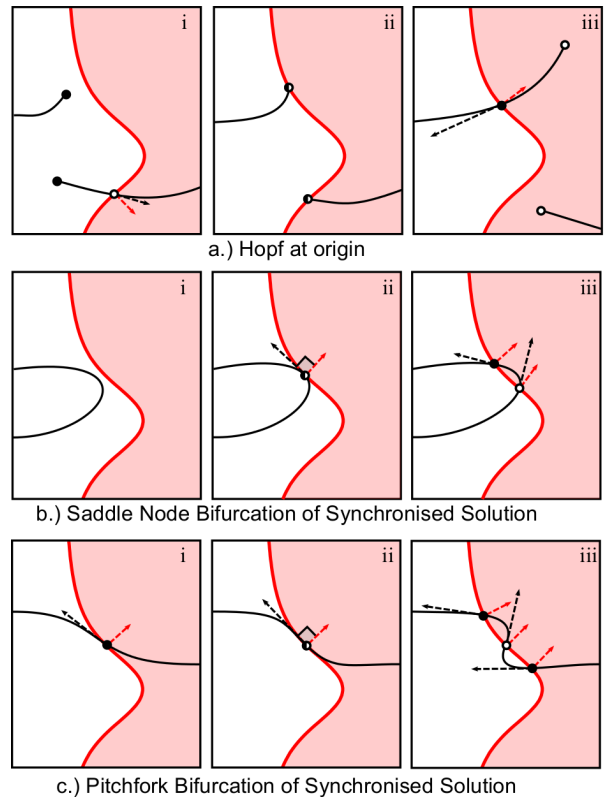}
 \caption{\label{fig:bifu} a) Hopf bifurcations of the trivial solution occur when $L$ is deformed such 
 that $F(0)$ passes from the set $U^c$ to $U$, creating(destroying) a new synchronised solution. In the topmost pane a), two systems
 are shown with stable trivial solutions solutions. When $L$ is deformed such that $F(0)$ is on the boundary $\partial U$, the trivial
 solution undergoes a supercritical(subcritical) Hopf bifurcation, creating(destroying) synchronised solutions as $F(0)$ passes into $U$.
 The middle pane b) is an example of how saddle-node bifurcations can occur in this system. 
 Suppose there is a convex segment of $L$ (i) and $L$ is deformed in such a way that this segment intersect $\partial U$ at one point (ii).
 At this intersection, both curves are tangential and thus a semi-stable synchronised solution is created. 
 As $L$ is further deformed, two synchronised solutions will be created, one stable and one unstable (iii).
 The bottom pane c) shows how a synchronised solution can bifurcate into three synchronised solutions 
 through a pitchfork bifurcation. A synchronised solution already exists (i) in the example system.
 When $L$ is deformed such that it intersects tangential,  the synchronised solution becomes semi-stable (ii).
 Continuing this deformation causes the synchronised solution to become unstable(stable) as two new stable(unstable) synchronised 
 solutions appear through a supercritical(subcritical) pitchfork bifurcation of periodic orbits.}
\end{figure}

In this framework, saddle-node and pitchfork bifurcations of periodic orbits occur when 
\[
\ip{\D{F}{r}\bigg|_r}{\D{}{\mu}\left(\frac{1}{f(\mu)}\right)\bigg|_{1+\I\Omega}}  = 0,
\]
that is when $F$ and $\partial U$ intersect tangentially. 
One can recover this condition by considering Eq.~\eqref{eqev} and requiring that both $\phi_d(0)=0$ and $\iD{\phi_d}{\lambda}(0) = 0$. 
To see how this can occur geometrically
suppose we have a segment of $F$ in $U^c$ which we continuously deform until the bifurcation point at which it just touches 
$\partial U$. 
At this point of contact $F$ and $\partial U$ will create a synchronised solution with a two dimensional centre manifold. 
Further deformation of $F$ would then result in a greater or lesser number of synchronised solutions similar to one dimensional
bifurcation theory.
Figure~\ref{fig:bifu} shows saddle-node and pitchfork bifurcations of synchronised solutions.

With the possibility of complex solutions to Eq.~\eqref{eqev} 
it is possible that the system could undergo a Hopf-bifurcation to a 
torus. 
However, if we suppose the conjecture in the previous section holds; that for unimodal distributions the largest eigenvalue is real, 
any Hopf bifurcation could not result in stable toroidal motion and thus are not of interest.
As we will see in the following section, this conjecture breaks down when we consider multi-modal distributions.

\section{Bimodal Distributions}\label{sec:bm}
One of the key results of \cite{cathy2012} was the existence of in-phase an 
anti-phase synchronised solutions for a population of oscillators with two 
distinct frequencies. Using the results in section \ref{sec:fp}, these appear as 
high frequency (in-phase) and low frequency (anti-phase) solutions.
\subsection{Synchronised Solutions}
To show this, we consider a bi-modal population of oscillators consisting of two 
Cauchy distributions 
\[
 g(\omega) = \frac{\sigma}{2\pi}\left[\frac{1}{\sigma^2 + 
(\omega-\Delta\omega)^2}+\frac{1}{\sigma^2 + (\omega + \Delta\omega)^2} \right],
\]
with identical spread $\sigma$ and modal values at $\pm \Delta\omega$. 
In the limit as $\sigma\rightarrow 0$ we recover the two oscillation problem previously investigated. 
$f(\mu)$ has a relatively simple closed form expression
 \[
 f(\mu) = \left[\mu + \sigma +\frac{\Delta\omega^2}{\mu+\sigma}\right]^{-1},
\]
which allows us to apply our main results Eq.~\eqref{eqfp} and show that synchronous solutions 
satisfy
$ F(r) = 1+\sigma+\I\Omega+(\Delta\omega^2)/(1+\sigma+\I\Omega)$.
We can eliminate $\sigma$ from the equation by rescaling the system, 
$\Delta\omega \rightarrow \Delta\omega/(1+\sigma)$, $\Omega 
\rightarrow\Omega/(1+\sigma)$ and the coupling function $F(r) \rightarrow 
F(r)/(1+\sigma)$. 
Essentially, this shows that in the continuum limit, the bimodal Cauchy with equal spread
is identical to a scaled version of the two oscillator problem. By increasing the spread $\sigma$, 
the effective coupling strength $\mod{F(r)}$ and frequency spread $\Delta\omega$ will be reduced. 
Whilst the scaled system may have different dynamic states, we can follow the same 
procedure to determine the mean field frequency $\Omega$ and amplitude $r$ which (in the rescaled system) satisfy
\begin{equation}
 F(r) = 1+\I\Omega+\frac{\Delta\omega^2}{1+\I\Omega} \label{bm:fp}.
\end{equation}
Separating real and imaginary parts, and performing some algebraic manipulation to eliminate $\Delta\omega^2$ allows us to express the
mean field frequency $\Omega$ in terms of the mean field amplitude $r$, 
\begin{subequations}
\begin{equation}
 \Omega =\frac{F_i(r)}{2- F_r(r)}.\label{bm:omega}
\end{equation}
Which allows us to express $\Delta\omega$ 
 \begin{equation}
   \Delta\omega^2 = \mod{F(r) - 2}^2 \frac{F_r(r) -1}{(F_r(r) - 
2)^2}.\label{bm:r}
 \end{equation}
\end{subequations}
as a function of $r$.
We define state $(r,\Omega)$ occurring when $r$ approaches the 
solution to $F_r(r) = 2$ with $|F_i(r)|>0$ as the in-phase state. This state is characterised by a large mean field frequency $\mod{\Omega} \gg 1$. 
For $\Omega$ positive in this limit almost all $\omega \ll \Omega$ and so, referring to 
Eq.~\eqref{eqfp} and figure~\ref{fig:fixedpoint}, almost all $z_j$ are close to the origin and on the lower arc, or clockwise side of
the quarter of the circle centred at $rF(r)/2$. 
Similarly for $\Omega$ negative, almost all $z_j$ will be on the upper arc, or anti-clockwise side of
the quarter of the circle in figure~\ref{fig:fixedpoint}.
In either case if the radius of the circle is big enough, the individual oscillators will appear as a cluster of fixed 
points with essentially the same phase and amplitude, hence the `in-phase' state.

We can similarly characterise the low frequency state as the anti-phase state as occurring when 
$-\Delta\omega <\Omega <\Delta\omega$. In terms of the density in 
Figure~\ref{fixedpoint}, this places the modal values on opposing arcs of the 
circle of fixed points. Oscillators $z_j$ with $\omega_j\approx \Delta\omega$ 
are fixed on the upper arc and are close to the origin when $\omega_j > \Omega 
+1$. Similarly, oscillators $z_j$ with $\omega_j \approx -\Delta\omega$ are on 
the lower arc and are close to the origin when $\omega_j< \Omega-1$.
If $\Omega \approx 0$ these solutions will appear symmetric, and if $rF(r)$ is 
big enough, they will appear approximately out of phase.
As $\Omega$ moves away from but stays between $(-\Delta\omega,\Delta\omega)$ the 
solutions become asymmetrical, with $\Omega$ closer to the positive mode 
$|\Omega-\Delta\omega| < |\Omega+\Delta\omega|$, producing larger amplitude 
solutions for oscillators with positive natural frequencies.
\begin{figure*}
\includegraphics{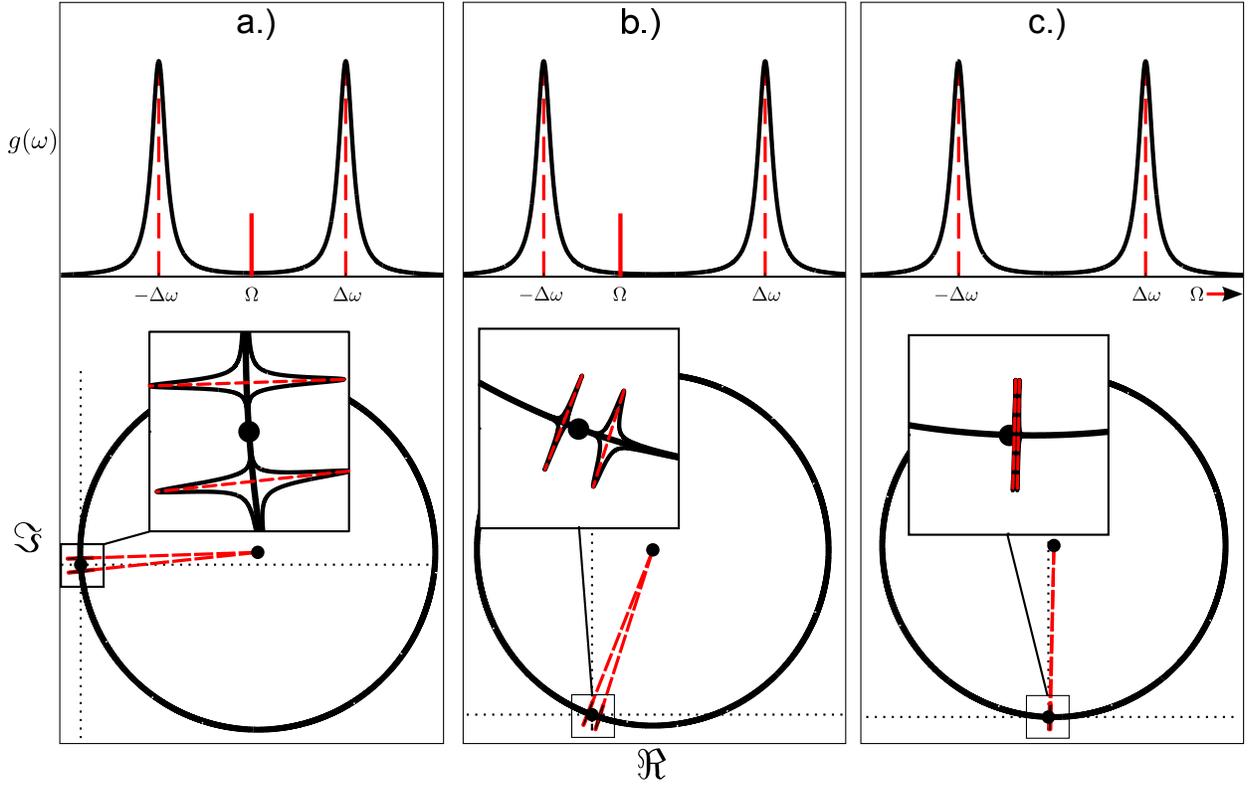}
 \caption{\label{fig:soln}
In phase and anti-phase solutions arise from differing values of the mean field 
rotational velocity $\Omega$. Both a.) and b.) are examples of anti-phase 
solutions which occur when $\Omega$ is between the modal values $-\Delta\omega < 
\Omega <\Delta\omega$. In a.), $\Omega \approx 0$ gives 
solutions where oscillators with frequencies $\omega_i<0$ appear on the lower 
arc and oscillators with $\omega_i>0$ appear on the upper arc. In this way the 
density on the circle inherits the symmetry of the distribution of frequencies 
and produces a symmetric, anti-phase solution. 
Solutions with slightly larger mean field frequency $\mod{\Delta\omega} \ll 
\mod{\Omega}>0$ map a larger portion of oscillators to one arc as represented in b.). Oscillators with natural 
frequency closer to $\Omega$ will be positioned further from $0$ on the circle 
and thus will have larger amplitude oscillations which produces the asymmetric 
anti-phase solution. In phase solutions c.) occur when
$\Omega$ is larger than the largest oscillator frequency so that all oscillators 
are on the same arc. If $\Omega$ is sufficiently large, the oscillators are 
packed closer to the origin and have similar amplitudes.}
\end{figure*}
\subsection{Stability}
\subsubsection{Amplitude Death Solution}
We now turn to the stability of the dynamic states. Again we 
can scale out $\sigma$  by setting $\lambda = \lambda/(1+\sigma)$ in Eq.~\ref{eq:zeroev} and~\ref{eqev}.

For the critical point at the origin, the stability is determined by 
Eq.~\eqref{eq:zeroev}
\[
 F(0)= 1+\lambda+\frac{\Delta\omega^2}{1+\lambda}
\]
which has eigenvalues satisfying
\[
 \lambda^2 + \lambda[2 - F(0)] + 1 - F(0) +\Delta\omega^2 = 0.
\]
As this is a complex quadratic equation of the form $\lambda^2 + a_1\lambda + a_0=0$, all the zeros will be
 in the left half plane if and only if
$ \Re{a_1}>0$, $\Re{a_1} \Re{(a_1\conj{a_0})} -(\Im {a_0})^2> 0$.
The first condition requires
\[
 2 >F_r(0),
\]
i.e. that there is insufficient forcing to overcome the systems inherent stability.
The second condition requires 
\begin{equation}
    \Delta\omega^2 > \mod{F(0) - 2}^2 \frac{F_r(0) -1}{(F_r(0) - 
2)^2} \label{bm:hopf0}
\end{equation}
which states that there must be a $r=0$ solution to Eq.\eqref{bm:r}.
Furthermore, if we vary $\Delta\omega$ so that there is equality in Eq.~\eqref{bm:hopf0} then a Hopf bifurcation occurs.
\subsubsection{Synchronised Solution}
\begin{widetext}
For the synchronised solution $F(r) = 1/f(1+\I\Omega)$, we note that
\begin{eqnarray*}
 \frac{1}{f(1+\I\Omega)} -\frac{1}{f(1+\I\Omega+\lambda)} &=&\frac{\lambda}{(1+\I\Omega)}\left[\frac{\Delta\omega^2}{(1+\I\Omega) +\lambda} -(1+\I\Omega)\right].
\end{eqnarray*}
Substituting this into Eq.\eqref{eqev} and looking the the solutions where $\Re[\lambda]\ne-1$, produces a real valued quartic of the form
$\lambda q(\lambda)$ where, $q(\lambda) = a_3\lambda^3 + a_2\lambda^2+a_1\lambda +a_0$ and 
\begin{eqnarray}
a_3 &=& \mod{\kappa}^2  \qquad \kappa = 1+\I\Omega\\
 a_2 &=& 2(\mod{\kappa}^2-\Delta\omega^2)-\mod{\kappa}^2\Re\left[r\D{F}{r}\right]\label{a2}\\
 a_1 &=&\mod{\Delta\omega^2 -\kappa^2}^2 + \Delta\omega^2\Re\left[\kappa r\D{F}{r}\right] - 2|\kappa|^2\Re\left[r\D{F}{r}\right]\label{a1}\\
 a_0&=&\Delta\omega^2\Re\left[\kappa^2r\D{F}{r}\right]-\Re\left[r\D{F}{r}\right]\mod{\kappa}^4\label{a0}.
\end{eqnarray}
\end{widetext}
Now, $\lambda q(\lambda)$ always has a zero eigenvalue solution; at the point of intersection in $F$ space, associated with the
rotating centre manifold. 
As $q$ is a cubic, there are either three real solutions, or one real and one conjugate pair. These solutions are 
in the left half plane if all coefficients are positive, and 
$ a_2a_1 > a_3a_0$. 
As we continuously change $\Delta\omega$ or deform $F$, a saddle-node bifurcation occurs as (one of) the real eigenvalue(s) pass through zero, 
which occurs when $a_0=0$ and all other terms are positive.
We can also identify Hopf bifurcations, which occur when $a_2a_1 = a_3a_0$ and $a_0/a_2>0$.

\begin{figure}
 \includegraphics{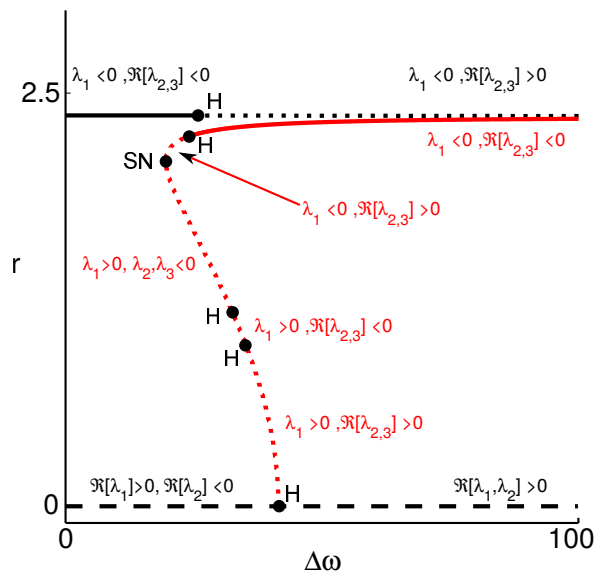} 
 \caption{\label{fig:dwfp}Synchronised solutions for the mean field $r$ as a 
function of $\Delta\omega$  in the parameter regime $(\delta,\epsilon)_2$, equivalent to Figure 
7,\cite{cathy2012}. For small values of $\Delta\omega$, the system has a stable, 
large $\Omega$, in-phase synchronised solution and an unstable origin. As 
$\Delta\omega$ increases, a pair of unstable solutions with $0<\mod{\Omega}< 
\mod{\Delta\omega}$ are created via a saddle-node bifurcation. 
As $\Delta\omega$ increases, one branch of the saddle-node 
gains stability via a Hopf bifurcation. The other branch collides with the 
origin in a subcritical Hopf bifurcation. 
The signs of the eigenvalues along each branch show how the dimension and nature of the unstable manifold
changes as $\Delta\omega$ changes}
\end{figure}

Figure~\ref{fig:dwfp} plots the location of the fixed points $r$ and the 
bifurcations as a function of $\Delta\omega$ for the system considered in 
\cite{cathy2012}. For $\Delta\omega$ small, this system has a stable in-phase 
synchronised solution and and unstable critical point at the origin. 
The in-phase solution loses its stability via a Hopf bifurcation further along in $\Delta\omega$.
As $\Delta\omega$ increases, a pair of unstable anti-phase solutions are created via 
a saddle-node bifurcation. One of these solutions gains stability via a 
Hopf bifurcation of periodic orbits. The other branch of the saddle-node bifurcation persists 
for a short range of values in $\Delta\omega$, undergoing a Hopf bifurcation which may create
quasi-periodic solutions, before colliding with the origin in a subcritical Hopf bifurcation.

From the applications point of view, we are predominantly interested in transitions from stable synchronised motion to 
unstable behaviour which may be of use as a switching or detection mechanism, however it is
important to note that a variety of bifurcations may occur which do not change the overall stability characteristics
of a synchronised solution. Indeed, the existence of Hopf bifurcations along an unstable branch of solutions suggests 
the existence of stable toroidal or quasi-periodic solutions which have been seen to occur in coupled oscillator systems
\cite{matthews91}. In the case of the two oscillator system\cite{cathy2012}, toroidal solutions were seen, but only existed
for a very small window in parameter space.

\section{Discussion and Conclusions}\label{sec:conc}

In this paper we have shown the existence and stability of collective behaviour 
in a system of identically damped harmonic oscillators with nonlinear all-to-all coupling and different natural frequencies.
The model, which to our knowledge has not been covered by previous results on collective behaviour, arises from
recent advances in optomechanics. 
We have been particularly interested in the existence and stability
of synchronised states, where every individual oscillator rotates at the same frequency
and in how changes in the coupling can cause the synchronised state to lose stability in favour of the trivial state.
We find that the existence and stability of these states is due to a play-off between the nonlinear coupling and the distribution 
of natural frequencies. 

In the synchronised state, the phase and amplitude of particular oscillators
were shown to lie on a circle co-rotating with the mean field; the arithmetic mean of the
oscillators complex co-ordinates. This result is independent of the distribution of the natural frequencies of the 
oscillators, however the position of oscillators on this circle does depend on the distribution. Significantly it was 
shown that the synchronised state is identified by a 
unique pair comprising of the mean field amplitude, $r$, and the mean field rotational velocity, $\Omega$. Indeed if one knows such
values in a synchronised state, one can back out the phase and amplitude of each individual oscillator using Eq.~\eqref{fixedpoint}.

The existence condition for $(r,\Omega)$ captures the 
play off between the nonlinear coupling function $F(r)$ and the distribution via the dispersion function, $f(\mu)$.
The stability condition for that solution, which is also a function of $(r,\Omega)$, additionally depends on the
derivative of the nonlinear coupling function.

As seems to be common in coupled oscillator problems, choosing Cauchy distributed frequencies drastically simplifies
the problem; to the point where the existence and stability of the synchronised states only
depend on the mean field amplitude, $r$. This implied radial symmetry simplifies the kinds of bifurcations that can occur. 
We found that the trivial state can lose (or gain) stability
via a Hopf bifurcation, and that stable synchronised states can be created and destroyed via saddle node or pitchfork bifurcations
of limit cycles.

In general unimodal distributions, such as the Gaussian and Uniform distributions, the existence, stability and bifurcations of the
synchronised state depend on both $r$ and $\Omega$.
It is then useful to develop a geometric method to determine the existence, 
stability and bifurcations of the synchronised states. This was made possible by the fact that the unimodal assumption guarantees that the dispersion function 
is a conformal map (see appendix .\ref{appendix3}). 
This geometric method allows us to contrast the effect of different unimodal distributions and explain why, for a particular 
system, a Guassian frequency distribution may produce stable synchronised states while for the same system
a Cauchy frequency distribution has only stable trivial solutions.
However the geometric method also suggests that the range of bifurcations that can occur in unimodal distributions does not depend 
on the type of distribution and that certain bifurcations that can occur in bimodal systems, such as Hopf bifurcations
to a stable torus, do not occur in systems with unimodal frequency distributions.

Finally we considered a bimodal frequency distributions that provides a means to understand in-phase and anti-phase 
solutions found previously\cite{cathy2012}. In fact in a suitable limit the distribution models two coupled oscillators.
Our results show that the in-phase state
appears as solutions with high frequency oscillations of the mean field. Low frequency solutions produce symmetric and asymmetric 
anti-phase behaviour. Further we established a relationship between the separation of
peaks in the frequency distribution and the existence and stability of both in-phase and anti-phase 
synchronised solutions for a physically relevant coupling function.

\appendix
\section{A coupling function from optomechanics}
\label{model}
We provide a brief review of the motivating physical system, as derived 
in\cite{cathy2012}, which describes the semi-classical equations motion of a 
collection of nano-mechanical resonators linearly coupled to an optical cavity 
which is driven by a single optical mode. The vibrational mode $x_i$ of each 
resonator satisfies
\begin{equation}
x_i'' + \omega_0^2x_i  = -2\gamma\left(x_i' +\omega_0\omega_ix_i 
+\frac{\omega_0}{4}\mod{\alpha}^2\right) \label{eq:resde}
\end{equation}
where $\alpha$ is the cavity amplitude and $\omega_0$ is the average natural 
frequency of the resonators and $\gamma$ is the coefficient of damping, consider 
small. We also assume the the effect of the coupling and the variation in 
natural frequency are also small; on the order of $\gamma$. The variation in 
natural frequency $\omega_i$ is assumed to be sampled from some distribution 
$g(\omega)$.

The problem can be simplified by applying the method of multiple scales 
\cite{glendinning} or 'two-timing', which separates the fast oscillation from 
slowly varying phase and amplitude dynamics. 
Defining the slow time scale $\tau =\gamma t$ we assume solutions to 
Eq.~\eqref{eq:resde} are of the form 
$x(t) = x_i^{0}(t,\tau) + \gamma x_i^{1}(t,\tau) + O(\gamma^2)$, where 
$x_i^{(0)}(t,\tau) = z_i(\tau)\exp(\I\omega_0t) + 
z_i^\dagger(\tau)\exp(-\I\omega_0t)$ and $x_i^{(1)}(t)$ is bounded. Also make 
the assumption that the cavity forcing $\I\mod{\alpha}^2/4$ has a Fourier 
expansion in the form
\[
 \I\mod{\alpha}^2/4 = \sum_{m\in \mathbb{Z}}\mathfrak{f}_m(\tau)\e^{m\omega_0t} 
+O(\gamma).
\]
This produces the complex amplitude equations, previously found in 
\cite{cathy2012}
\begin{equation}
\D{z_i}{\tau} = -(1-\I\omega_i)z_i + \mathfrak{f}_{1}(\tau) \label{eqds}.
\end{equation}
 where $\mathfrak{f}_1$ is the first rotating term in the formal Fourier 
series expansion of $\I\mod{\alpha}^2/4$ with basis $\exp(\I m\omega_0t)$. 

The dynamics of the optical cavity are described by a forced, damped and 
parametrically detuned complex oscillator equation
\begin{equation}
\alpha' = -(1+\I\delta)\alpha - \I \alpha\sum_{i=1}^n\frac{x_i}{n} 
-\I\epsilon\label{eqcav}
\end{equation}
where $\delta$ is the difference between the natural frequency of the cavity and 
the external forcing frequency and $\epsilon$ is the forcing amplitude. 
Eq.\eqref{eqcav} can be expressed in terms of a mean field $x(t) = 
2\mod{z}\cos(\omega_0t + \Theta)$ where
\[x(t) = z\e^{\I\omega_0t} +z^\dagger\e^{-\I\omega_0t}, \quad z = 
\frac{1}{n}\sum z_i, \] 
and can be integrated by using the integrating factor $\exp[(1+\I\delta)t + 
\I2r\sin(\omega_0t +\Theta)/\omega_0]$ in terms of field.
Fortunately, the solution is automatically a Fourier series in terms of $\exp(\I 
m\omega_0t)$ and so we define the coupling term $\mathfrak{f}_1(\tau) = zF(|z|)$ 
and the nonlinear coupling function 
\begin{equation}
                 F(\mod{z}) = \frac{1}{\mod{z}}\sum_{m=0}^\infty 
c(m)J_m\left(\frac{2\mod{z}}{\omega_0}\right)J_{m+1}\left(\frac{2\mod{z}}{
\omega_0}\right)\label{eqcoup}
                \end{equation}
with complex coefficients given by their real and imaginary parts
\begin{widetext}
\begin{eqnarray*}
c_r(m)&=&\frac{\epsilon^2\omega_0}{4}\bigg[\frac{1}{1+(\delta+m\omega_0)^2}
\times\frac{1}{1+(\delta + (m+1)\omega_0)^2} 
-\frac{1}{1+(\delta-m\omega_0)^2}\times\frac{1}{1+(\delta - (m+1)\omega_0)^2} 
\bigg]\label{eqcmr}\\
  c_i(m)&=&\frac{\epsilon^2}{4(4+\omega_0^2)}\bigg[\frac{\omega_0(\delta 
+m\omega_0)+2}{1+(\delta+m\omega_0)^2}-\frac{\omega_0(\delta + (m+1)\omega_0) 
-2}{1+(\delta+(m+1)\omega_0)^2}+\frac{\omega_0(\delta-m\omega_0) 
-2}{1+(\delta-m\omega_0)^2} -\frac{\omega_0(\delta - (m+1)\omega_0) 
+2}{1+(\delta-(m+1)\omega_0)^2}\bigg].
\end{eqnarray*}
\end{widetext}
The constant terms of $F(\mod{z})$ will produce the radiation induced damping 
and frequency pulling found in \cite{Seok2012}. In terms of the detuning 
$\delta$ the shape of all $c_r(m)$ and $c_i(m)$ are Cauchy like or 
dispersion($\frac{x}{1+x^2}$) like with peaks and zero crossing at $\delta = \pm 
m\omega_0$ and $\delta = \pm (m+1)\omega_0$.

Throughout this work we have used the same parameter set as\cite{cathy2012}, 
with $\omega_0=2$ and 
$\gamma = 0.0001$. The parameters related to physically controllable quantities 
$\epsilon$ and $\delta$ are experimentally achievable \cite{cathy2012} affect 
the magnitude ($|F|~ O(\epsilon^2)$) and shape of $F(\mod{z})$.
In particular, our model has been normalised such that 
$\varepsilon\sqrt{NG/\gamma} = \epsilon$ and hence $\varepsilon\sqrt{NG/\gamma} 
\approx 1$ and $\epsilon\approx 100$ describe the same system. 

For sections~\ref{sec:lor} and~\ref{sec:uni} three parameter sets have been used as examples: 
\begin{eqnarray}
 (\delta,\epsilon)_1 &=&(-1.5,70)\\
 (\delta,\epsilon)_2 &=& (-1.5,150)\\
 (\delta,\epsilon)_3 &=& (-4.5,15)
\end{eqnarray}
Computations have been performed by truncating Eq.~\eqref{eqcoup} at $20$ terms.

For the bimodal section we used $\delta = -1.5$, $\epsilon = 100$. The calculations were performed 
using an $4$ term Taylor series expansion as opposed to directly expanding the sum.
This particular expansion was chosen to conform with the results in\cite{cathy2012}.

\section{Derivation of Characteristic function}
\label{appendix2}
Let $a,b$ be complex numbers, $G$ be an $n\times n$ diagonal matrix with entries $G_{jj} = \I\omega_j - \mu$ and $U$ be an $n\times n$ matrix with ones in every entry.
Thus we define
\[
 J = \left(\begin{matrix}
      G +  aU & bU\\
      \conj{b}U & \conj{G} +  \conj{a}U
     \end{matrix}\right)
\]
and seek the characteristic equation for $J$.
Firstly we will consider the case when $b=0$. It is sufficient to look at the eigenvalues of $J_{11}=G +  aU$ which was 
investigated by Mirollo and Strogatz in\cite{Mirollo90}, where they show that $J_{11}$ has a characteristic equation in terms of the finite dispersion function \eqref{fmun};
\[
 \phi_{11}(\lambda) =\left[1- af_n(\mu+\lambda)\right] \prod_{j=1}^n(\lambda +\mu - \I\omega_j).
\]

So define 
\[
 \phi_1(\lambda) =\det\left[(J_{11}-I\lambda)(J_{22}-I\lambda)\right] = \phi_{11}(\lambda)\conj{\phi}_{11}(\lambda)
\]
\begin{widetext}

Suppose $b\ne0$ and $J_{11} - I\lambda $ is invertible (and hence $\conj{J_{11}} -I\lambda$ is invertible).
Then by factorisation,
\begin{eqnarray*}
\det(J-I\lambda) &=& \det[J_{11}-I\lambda]\det[\conj{J_{11}} -I\lambda]\\
& &\times \det\left[I-  |ab|^2 \left(\frac{G -I\lambda}{a}+  U \right)^{-1}U\left(\frac{\conj{G} -I\lambda}{\conj{a}}+  U\right)^{-1}U\right].
\end{eqnarray*}
Now the first factor is nothing but $\phi_1(\lambda)$ as given above. For the second factor we write the matrix of ones as the outer product of two vectors of ones
$U = uu^T$, and apply the matrix determinant lemma to give
\[
 \det(J-I\lambda) = \phi_1(\lambda)\left[1-  \frac{|b|^2}{|a|^2}u^T \left(\frac{G -I\lambda}{a}+  U \right)^{-1}uu^T\left(\frac{\conj{G} -I\lambda}{\conj{a}}+  U\right)^{-1}u\right].
\]1
We use the fact that
\[
 u^T\left(\frac{G -I\lambda}{a}\right)^{-1}u = -af_n(\mu+\lambda)
\]
which, when combined with the Sherman-Morrison formula, shows 
\[
 u^T \left(\frac{G -I\lambda}{a}+  U \right)^{-1}u = \frac{af_n(\mu+\lambda)}{af_n(\mu+\lambda)-1}.
\]
The same thing can be done with the conjugate term, whilst noting that $f_n(\conj{\mu}+\lambda) = \conj{f}_n(\mu+\conj{\lambda})$, to produce
\[
 \det(J-I\lambda)= \phi_1(\lambda)\left[1 - |b|^2\frac{f_n(\mu+\lambda)\conj{f_n}(\mu+\conj{\lambda})}{(af_n(\mu+\lambda)-1)(\conj{af_n}(\mu+\conj{\lambda})-1)}\right].
\]
\end{widetext}
Expanding $\phi_1(\lambda)$ and assuming $f_n$ is not zero. Then $\det(J-I\lambda) = \phi_l(\lambda)\phi_d(\lambda)/(f_n(\mu+\lambda)\conj{f_n}(\mu+\conj{\lambda})$ where
\begin{eqnarray*}
\phi_l(\lambda) &=&  \prod_{j=1}^n((\lambda+\mu_r)^2 + (\mu_i-\omega_j)^2)\\
\phi_d(\lambda)&=& (f_n(\mu+\lambda)^{-1}-a)(\conj{f_n}(\mu+\conj{\lambda})^{-1} -\conj{a})  -\mod{b}^2
\end{eqnarray*}
 
\section{Properties of the dispersion function}
\label{appendix3}

\subsection{Proof of Lower bound on $\Re[1/f]$}
Consider for $\mu_r>0$
\[
\Re[f(\mu)]  = \int_{-\infty}^\infty \frac{\mu_r}{\mod{\mu - \I\omega}^2}\df{G(\omega)} 
\]
which implies
\[
\frac{\Re[f(\mu)]}{\mu_r} = \int_{-\infty}^\infty\frac{1}{\mod{\mu - \I\omega}^2}\df{G(\omega)}.
\]
However 
\[
 \mod{f(\mu)}^2 =\mod{\int_{-\infty}^\infty\frac{1}{\mu - \I\omega}\df{G(\omega)}}^2\le \int_{-\infty}^\infty\frac{1}{\mod{\mu - \I\omega}^2}\df{G(\omega)}
\]
thus for $\mu_r >0$
\[
 \frac{\Re[f(\mu)]}{\mu_r} \ge \mod{f(\mu)}^2 \implies \mu_r \le \frac{\Re[f(\mu)]}{|f|^2}.
\]
Which gives the required lower bound
\[
 \mu_r \le \Re\left[\frac{1}{f(\mu)}\right]
\]

\subsection{Proof that $f$ is injective on $H$}
In order to prove this result, we begin by expanding two properties of $f$ established in Theroem 2 of\cite{Mirollo90} 
for symmetric unimodal distributions $g(\omega)$.
It was proved that for real valued $\mu_r=\Re[\mu]$, $f(\mu_r)$ is real valued and strictly decreasing function of $\mu_r$
and hence is bijective on this domain.
In the course of proving that real value $\mu$ implies real value $f(\mu)$ the imaginary part 
\[
 \Im[f] = \int_0^\infty\frac{\omega}{\mu_r^2+\omega^2}[g(\omega+\mu_i) - g(\omega-\mu_i)]\df{\omega}
\]
was shown to be strictly negative for $\mu_i>0$. It is easy to see that this is both necessary and sufficient. 
Setting $\mu_i \rightarrow -\mu_i$ we also find that $\mu_i<0 \iff \Im[f(\mu)] >0$.
We can thus say that $f$ maps the open first quadrant $Q_1$ of the complex plane to a subset of the open fourth quadrant $Q_4$ and hence 
if $f:Q_1\rightarrow f(Q_1)\subset Q_4$ is injective, then it is injective on the right halfplane.

Let us pick some $\mu_1,\mu_2\in Q_1$ and consider 
\begin{eqnarray}
f(\mu_1) - f(\mu_2) &=& \int_{-\infty}^\infty \left(\frac{1}{\mu_1-\I\omega}-\frac{1}{\mu_2-\I\omega}\right)g(\omega)\df{\omega}\notag \\
&=& (\mu_2-\mu_1)\int_{-\infty}^\infty \frac{g(\omega)}{(\mu_1-\I\omega)(\mu_2-\I\omega)}\df{\omega}\notag\\
&=&(\mu_2-\mu_1)I(\mu_1,\mu_2).\label{eqint}
\end{eqnarray}
We wish to show that $f(\mu_1) =f(\mu_2)$ implies $\mu_1 = \mu_2$, so it is sufficient to show that the integral is never zero. 
Using the symmetry of $g$ we have;
\begin{eqnarray*}
 I(\mu_1,\mu_2)
  &=&\int_{0}^\infty \bigg[\frac{1}{(\mu_1-\I\omega)(\mu_2-\I\omega)}\\
  &&\qquad+ \frac{1}{(\mu_1+\I\omega)(\mu_2+\I\omega)}\bigg]g(\omega)\df{\omega}\\
  &=&-2\int_0^\infty \frac{\omega^2-\mu_1\mu_2}{(\mu_1^2+\omega^2)(\mu_2^2+\omega^2)} g(\omega)\df{\omega}
\end{eqnarray*}
We shall call the imaginary part of the above integrand $k(\omega)$; 
\begin{eqnarray*}
 k(\omega) &=& \Im\left[
 -2 \frac{(\omega^2-\mu_1\mu_2)(\omega^2+\conj{\mu_1}^2)(\omega^2+\conj{\mu_2}^2)}{\mod{(\mu_1^2+\omega^2)(\mu_2^2+\omega^2)}^2} 
 \right]\\
 &=&
 \frac{\rho\omega^4+\alpha\omega^2 + \beta}{\mod{(\mu_1^2+\omega^2)(\mu_2^2+\omega^2)}^2},
\end{eqnarray*}
where
\begin{eqnarray*}
 \rho &=& 2\Im\left[\mu_1\mu_2-\conj{\mu_1}^2- \conj{\mu_2}^2 \right],\\
 \alpha &=& 2\Im\left[  \mod{\mu_1}^2\conj{\mu_1}\mu_2+  \mod{\mu_2}^2\mu_1\conj{\mu_2}-\conj{\mu_1}^2\conj{\mu_2}^2\right],\\
 \beta &=& 2\Im\left[\mod{\mu_1}^2\mod{\mu_2}^2\conj{\mu_1}\conj{\mu_2}\right].
\end{eqnarray*}
Clearly $\beta <0$ for all $\mu_1,\mu_2 \in Q_1$, so $k(\omega)$ is not identical to zero. 
Furthermore, since  $\rho >0$ and the numerator of $k(\omega)$ is a quadratic in $\omega^2$, $k(\omega)$ has one simple root on
the positive real line.

To show this integral is never zero, consider the domain $\omega \in [0,\infty)$ on which $g$ is a decreasing function. We invoke the layer cake representation
 where we express the integral
 \[\int_0^\infty k(\omega) g(\omega)\df{\omega}
 = \int_0^\infty\int_0^\infty k(\omega)\mathbb{I}_{\{g(\omega)\ge x\}}\df{x} \df{\omega}
 \] as an integral over
 the super-level sets of $g$; the intervals $[0,s)$. 
 Thus it suffices to show that the contribution from each level set
\[ K(s) = \int_0^s k(\omega) \df{\omega} \]
never changes sign for all $s$. 

Since $k(\omega)$ changes sign from negative to positive, $K(s)$ begins negative and increases 
in magnitude until a critical value $s_0$; when $k(s_0) = 0$, then decreases in magnitude for the remainder of the positive
real line. Thus it remains to show that $|K(s)|$ decreases to a limiting value, and that value is zero.
Working back up we can show that
\[
 \lim_{s\rightarrow\infty}K(s)
 = \Im\left[\int_{-\infty}^\infty\frac{1}{(\mu_1-\I\omega)(\mu_2-\I\omega)}\df{\omega}\right].
\]
We can evaluate the integral using contour integration over the right half plane, by setting $z = \I\omega$
\[
 \int_{-\infty}^\infty\frac{1}{(\mu_1-\I\omega)(\mu_2-\I\omega)}\df{\omega}
 =
 \int_H\frac{\I}{(z-\mu_1)(z -\mu_2)}\df{z}
\]
where $H$ is the right half plane. Note that the integrand is $O(\mod{z}^{-2})$ as $z\rightarrow \infty$, and the residues 
are $\res_{z=\mu_1} = \I(\mu_1-\mu_2)^{-1}$ and $\res_{z=\mu_2} = \I(\mu_2-\mu_1)^{-1}$, which clearly sum to zero.
so
\[
 \int_{-\infty}^\infty\frac{1}{(\mu_1-\I\omega)(\mu_2-\I\omega)}\df{\omega} =0.
\]
Thus $\lim_{s\rightarrow\infty}K(s) = 0$ and thus never changes sign. 
This implies that the integral in Eq.~\eqref{eqint} is never zero for $\mu_1\ne \mu_2$ and that both
$f:Q_1 \rightarrow f(Q_4)$ and, by symmetry, $f:H\rightarrow f(H)$ is injective.
\subsection{Proof that $f$ is a Conformal Map}
Consider $f$ as given in Eq.\eqref{fmu}.
Clearly $f$ is analytic for $\mu_r>0$ as the kernel $(\mu-\I\omega)^{-1}$ is analytic for all $\mu_r>0$.
We have proved that $f$ is injective on the right half plane for symmetric unimodal $g$ so it remains to 
identify the set of points where
\begin{equation}
\D{f}{\mu} = -\int_{-\infty}^\infty \frac{g(\omega)}{(\mu-\I\omega)^2}\df{\omega} = 0.  \label{eq:adf}
\end{equation}

Suppose that $g$ is symmetric and non-increasing on $[0,\infty)$.
In this case it has been previously shown by Mirollo and Strogtaz\cite{Mirollo90}
that $f(\conj{\mu})=\conj{f(\mu)}$,
that $f(\mu_r)$ is strictly decreasing for $\mu_r>0$ and that $\lim_{\mu_r\rightarrow\infty}f(\mu_r) = 0$.
Hence for $\mu$ real valued $\iD{f}{\mu} \ne 0$.

Observe that the imaginary part of Eq.\eqref{eq:adf} is given by
\begin{eqnarray*}
 \Im\D{f}{\mu} &=& -\int_{-\infty}^\infty \frac{2\mu_r\omega}{(\mu_r^2 +\omega^2)^2}g(\omega+\mu_i)\df{\omega}\\
	    &=& -\int_{0}^\infty \frac{2\mu_r\omega}{(\mu_r^2 +\omega^2)^2}[g(\omega+\mu_i) -g(\mu_i-\omega)]\df{\omega}.
\end{eqnarray*}
However, for all $\mu_i,\omega\ge0$, $|\omega-\mu_i| \le |\omega+\mu_i|$, thus $g(\omega-\mu_i)\ge g(\omega+\mu_i)$ 
which means the integrand is non-positive, and zero if and only if $\mu$ is real. 
Therefore $\iD{f}{\mu}=0$ has no solutions in the right half plane, which implies $f$ is conformal on the right half plane.

We can contrast this against a prototypical multi-modal distribution consisting of two delta functions, by setting $g(\omega) = [\delta(\omega-\Delta\omega) +\delta(\omega+\Delta\omega)]/2$
and thus
\[
 \D{f}{\mu} = -\left[\frac{1}{2(\mu-\I\Delta\omega)^2}+\frac{1}{2(\mu+\I\Delta\omega)^2}\right].
\]
This has zero solutions when $\mu = \Delta\omega$, which is in the domain and thus $f$ fails to be conformal.

\bibliographystyle{plain}
\bibliography{chaos}
\end{document}